\documentclass[12pt,a4paper,final]{iopart}

\usepackage{iopams}  
\usepackage{graphicx}
\usepackage[breaklinks=true,colorlinks=true,linkcolor=blue,urlcolor=blue,citecolor=blue]{hyperref}

\usepackage{subcaption}
\begin{document}
	
\title[New methods for creating superoscillations]{New methods for creating superoscillations}
	
\author{Leilee Chojnacki$^{1,2}$ and Achim Kempf$^{1,2,3}$} 
\address{$^1$Perimeter Institute for Theoretical Physics}
\address{$^2$Department of Physics, University of Waterloo}
\address{$^3$Department of Applied Mathematics and Physics, University of Waterloo}
		
\begin{abstract}
Superoscillating functions, i.e., functions that locally oscillate at a rate faster than their highest Fourier component, are of interest for applications from fundamental physics to engineering. Here, we develop a new method which allows one to construct superoscillations of arbitrarily high frequency and arbitrarily long duration in a computationally efficient way. We also present a method for constructing non-singular Schr\"odinger potentials whose ground state is a superoscillating wave function.
\end{abstract}
	

\vspace{2pc}

\section{Introduction}
It used to be thought that if the Fourier decomposition of a function does not contain frequency components above some frequency $\Omega$ then the function cannot oscillate at a rate faster than $\Omega$. However, the finding of strange weak values \cite{AAV} in the context of quantum weak measurements showed that there exist bandlimited functions which, despite having a finite bandlimit $\Omega$, locally oscillate faster than the function's overall bandlimit. 

Indeed, subtle combinations of low frequency components can interfere to generate a function which, locally, i.e., on some finite interval, oscillates at an arbitrarily high frequency. This paradoxical behaviour comes at a cost: the amplitudes in the local high frequency stretch must be very small, which also explains their late discovery. Such functions are termed superoscillating, with the rapidly changing stretches referred to as superoscillations.

In spite of their typically very small amplitudes, superoscillations are of significant interest in various applied fields. For example, superoscillations have been proposed as a means by which to focus pulses into arbitrarily small spatial and temporal windows \cite{Wong-Temporal}. 
It has also been shown, for example, that superoscillations naturally arise in 2D optical speckle patterns, and that they can fill as much as a third of the total pattern \cite{Dennis-Speckles}. 
{Optical superoscillations can also be generated from superpositions of diffraction free waves \cite{Makris}.}
However, as always, the amplitudes of the superoscillations quickly become extremely small as one increases the frequency and number of the superoscillations. 

{Indeed, for practical applications, it is important, therefore, to keep track of the ratio of the small amplitudes in the superoscillating stretch and the largest overall amplitudes in the signal. This ratio, which we may call the dynamic range (a commonly used term in signal processing), is a good indicator of the number of orders of magnitude that the experimental setup will need to be able to control.}

The fact that any practical application of superoscillations must bear with the need to handle very large dynamic ranges is a concern but it is worth pointing out that even quite large dynamic ranges can be experimentally accessible for use with superoscillations, as was, e.g., pointed out by Zheludev \cite{Zheludev}: for example, regarding superoscillations in electromagnetic waves, it is straightforward to produce on the order of $10^{20}$ photons/sec coherently using a laser, while it is also possible to detect luminosities as low as individual photons. 

This fact is important also for potential applications of superoscillations in the field of superresolution, (for reviews, see, e.g., \cite{Zheludev,Lindberg}):
the conventional diffraction limit is on the order of one half of the wavelength used. This means that to observe living tissues and cells without exposing them to ionizing radiation, one is limited to a resolution on the order of hundreds of nanometres. This is insufficient to observe, for example, proteins. Prior approaches to superresolution, see, e.g., \cite{Zheludev}, relied, for example, on evanescent waves produced at the boundary between two media, which can contain Fourier components above the bandlimit of the propagating field. The evanescent waves decay exponentially with distance from the boundary, which implies that any superresolving apparatus (superlens) must be placed such that the object under measurement is within the electromagnetic near field of the boundary, which in practice means at subwavelength distance. 
Superoscillations, on the other hand, do not require a substrate medium, and are therefore a promising successor to evanescent waves for the purposes of overcoming resolution limits. Additionally, it has been noted by Berry \cite{Berry-Faster} that while evanescent waves actually possess larger than bandlimit Fourier components, superoscillations do not. 

An application of superresolution aided by superoscillations could be, for example, the detection and resolution of centimetre-sized subterranean objects (such as landmines), \cite{Kempf-Prain}. This could be achieved with electromagnetic waves that possess only wavelengths large enough to penetrate the ground but that also possess superoscillations of short enough wavelength to resolve small objects. 
This approach to superresolution may not work in all circumstances, because in some materials superoscillating electromagnetic waves will be resonantly absorbed even though those resonance frequencies are not present in the Fourier spectrum of the incident wave. However, fortunately, this then offers an opportunity to use superoscillatory waves to explore the dynamical timescales of absorption and relaxation in that material \cite{Kempf-Prain}.

In the context of quantum mechanics, it has been proposed that a superoscillating wave function can be viewed as being akin to a spring-loaded particle \cite{Kempf-Ferreira-Unusual}: the particle can acquire a large net momentum simply by passing through a slit that is aligned with the superoscillating region so that only the superoscillating part of the wave function, with its short wavelengths, passes through the slit. 
{In short, a designed superoscillation can be used to boost particles. This process can be studied in reverse also, as in \cite{Barnett-Berry-Superweak} in the context of optical superoscillations: by measuring the momentum transferred to a particle in the vicinity of the optical superoscillation, a measurement of the weak values of momenta of the optical pattern is performed.}

Also the question of the persistence of superoscillatory behavior of wave functions under time evolution has been addressed, see e.g. \cite{Buniy-Colombo}, \cite{Aharonov-Colombo}, \cite{Aharonov-Evolution}.

It has also been shown how the framework of superoscillations allows for the steps made in a quantum random walk to have arbitrary size \cite{Quantum-Paradoxes}: by being in a superposition of modes representing bounded displacements, there is an exponentially suppressed probability for these modes to produce an arbitrary displacement. This is of interest in the field of quantum information, since it is thought that quantum random walks could lead to a universal quantum computer \cite{Aharonov2014}. 

While there is, therefore, a whole range of potential applications of superoscillations, the conventional methods for mathematically constructing superoscillatory functions - as discussed, e.g., in  
\cite{Makris}\cite{Kempf-Ferreira-Unusual}\cite{Lee-Ferreira-1} - are generally limited by numerical difficulties. 
{These methods typically involve two key procedures: first, the prescription of a set of points describing subwavelength features through which a bandlimited function must pass; second, the imposition of additional constraint(s) or assumptions to identify the desired superoscillatory function.}
For example, the method described in \cite{Kempf-Ferreira-Unusual} yields superoscillatory functions which possess the best possible dynamic range but this method requires the inversion of an ill-conditioned matrix whose condition number gets exceedingly worse as the frequency and number of superoscillations is increased. This makes it computationally prohibitively difficult to calculate functions with more than about 20 superoscillations whose frequency is twice the bandlimit. 
{Some methods to improve the conventional approaches have been suggested in \cite{Lee-Ferreira} \cite{Lee-Ferreira-2} by relaxing a variety of constraints in order to obtain a better conditioned matrix. Periodic superoscillations and particular constraint strategies are discussed in \cite{Katzav-Schwartz} and \cite{Tang-Garg-Kempf}. }

In this paper, in Sec.\ref{sec:multiplic}, we develop a new approach to the calculation of superoscillations. It overcomes these difficulties and allows one to straightforwardly and explicitly calculate superoscillatory functions with superoscillatory stretches of any length and frequency. We also calculate the scaling of the dynamic range of the so-obtained superoscillatory functions and find that it scales optimally. 

In Sec.\ref{sec:wavefunction} we then develop a method that may be useful for creating superoscillatory wave forms in quantum mechanical wave functions. To this end, we propose a new method for constructing a quantum Hamiltonian with a non-singular potential such that the ground state is any predetermined periodic superoscillating wave function.

\section{Generating superoscillations multiplicatively}
\label{sec:multiplic}
Our aim is to develop a method for generating superoscillations which allows one to directly and explicitly determine the functional form of a superoscillating signal without significant numerical costs. The strategy is to enforce the superoscillatory behavior by ensuring that the to-be-designed function possess prescribed close-by zeroes. To this end, the idea is to multiply {a finite number of} sufficiently low bandwidth functions that each contribute one or several zeros to the product. 
Concretely, consider a function of the form
\begin{equation}
S_N(t) = \prod_{i=1}^N b_i(t)
\end{equation}
where the factors $b_i(t)$ are functions with bandlimit $\Omega_i$. The bandlimit $\Omega$ of their product, $S_N(t)$ is: 
\begin{equation}
\Omega = \sum_{i=1}^N \Omega_i
\end{equation}
{\noindent To see this, consider two bandlimited functions $f(t)$ and $g(t)$, with respective bandlimits $\Omega$ and $\Omega^\prime$:}
\begin{equation}
\label{eq:FT of f}
f(t) = \frac{1}{\sqrt{2\pi}}\int_{-\Omega}^{\Omega} \tilde{f}(\omega) e^{i \omega t} d\omega~~,~~~~~~
g(t) = \frac{1}{\sqrt{2\pi}}\int_{-\Omega'}^{\Omega'} \tilde{g}(\omega) e^{i \omega t} d\omega'
\end{equation}
Their product, $h(t) := f(t)g(t)$, has a Fourier transform, $\tilde{h}(\omega)$, which obeys: 
\begin{eqnarray}
\tilde{h}(\omega) = \frac{1}{\sqrt{2\pi}}\int_{-\infty}^{\infty} h(t) e^{-i \omega t} dt = \frac{1}{\sqrt{2\pi}}\int_{-\infty}^{\infty} f(t)g(t) e^{-i \omega t} dt\\
 = \frac{1}{\sqrt{2\pi}^3}\int_{-\infty}^{\infty}\int_{-\Omega}^{\Omega} \int_{-\Omega^\prime}^{\Omega^\prime} \tilde{f}(\lambda)e^{i \lambda t} \tilde{g}(\lambda^\prime)e^{i \lambda^\prime t} e^{-i \omega t} dt d\lambda d\lambda^\prime \\
 = \frac{1}{\sqrt{2\pi}}\int_{-\Omega}^{\Omega} \int_{-\Omega^\prime}^{\Omega^\prime} \tilde{f}(\lambda)\tilde{g}(\lambda^\prime) \delta(\lambda+\lambda^\prime - \omega) d\lambda d\lambda^\prime
\end{eqnarray}
We notice that $\tilde{h}(\omega)$ vanishes for any value of $\omega$ outside of the range $[-(\Omega+\Omega^\prime),\Omega+\Omega^\prime]$, i.e., the product function $h$ has the bandlimit $\Omega+\Omega'$. 

When generating an $\Omega$-bandlimited function $S_N(t)$ by multiplying $\Omega_i$-bandlimited factor functions $b_i(t)$, the locations of the zeroes in $S_N(t)$ arise cumulatively from the zeroes of each of the factor functions $b_i(t)$. By using a suitable product of factor functions, one can obtain an $\Omega$-bandlimited function $S_N(t)$ which possesses zeroes that are arbitrarily closely spaced, and which is therefore superoscillating. 
{Adequate control of the phase of each factor function is, therefore, essential to insure that the zeroes are placed at the desired locations.}

{We conclude that by multiplying a finite number, $N$, of bandlimited functions with suitably close zeros, we can generate at least $(N-1)/2$ superoscillations, given that each factor function contributes at least one zero, as we illustrate with examples in the next two subsections.}

With the new method, it is intuitively clear that the dynamic range of the so-obtained superoscillating functions scales exponentially with the number of superoscillations and polynomially with the frequency of the superoscillations, which we will also show rigorously. This scaling behavior matches the scaling behavior which was proved in \cite{Kempf-Ferreira-Unusual} for the superoscillations with the optimal dynamic range. This means that the gain in computational efficiency of the new method comes with no significant deterioration of the dynamic range of the superoscillatory waves that it produces.     
The new method can, of course, be used to construct both square integrable or periodic superoscillating functions and we will study both cases. 

\subsection{Generation of periodic superoscillating functions}
As examples of periodic factors, let us use the bandlimited trigonometric functions sine or cosine. We may generate a periodic superoscillating function, for example, by multiplying a sinusoid by any number of copies of itself translated by a set of arbitrarily small displacements $\epsilon_i$, such as
\begin{equation}
S(t) = \prod_{n=1}^{N} \sin\left(\frac{\Omega}{N}(t-\epsilon_n)\right)
\end{equation}
where the displacements $\epsilon_i$ will determine the spacing between the zeroes of the function. Once this spacing is less than one half the shortest wavelength corresponding to the overall bandlimit, the function will be superoscillating. Each pair of zeroes spaced in this way will form the nodes of one half oscillation. 

We may also parametrize the above construction in terms of the desired local superoscillating frequencies $\omega_i=2\pi\nu_i = \pi/\epsilon_i$. 
In this example, each sine function has the same individual bandlimit $\Omega_i = \Omega/N$, where $\Omega$ is the overall bandlimit.

These constructions can for example be made antisymmetric about the origin for an odd number of superoscillations $N$
\begin{equation}
S(t) = \sin\left(\frac{\Omega}{N}t\right)\prod_{k=-1,1}\prod_{n=1}^{\frac{N-1}{2}} \sin\left(\frac{\Omega}{N}(t+k\epsilon_n)\right)
\end{equation}
from which even constructions can be obtained by, for example, squaring. Note that the factor functions in this case must have individual bandlimits $\Omega/2N$.

\begin{figure}[!t]
	\begin{center}
		\includegraphics[width=0.8\columnwidth]{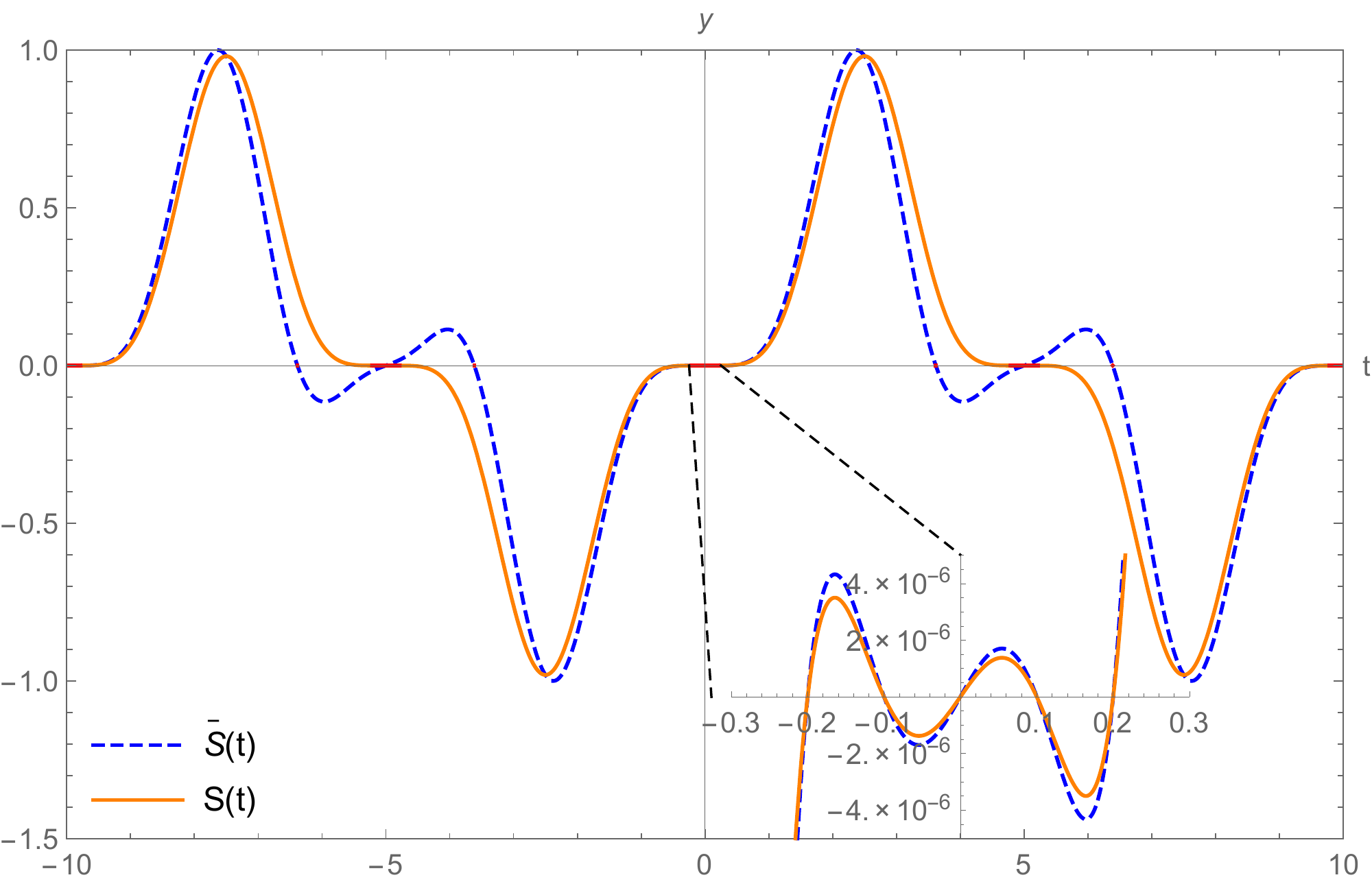}
		\caption{\label{fig:Old_Sin_comparison}
			Two periodic superoscillating curves $\bar{S}(t)$ and $S(t)$ obtained by the additive and multiplicative approaches respectively. The superoscillations are displayed in the inset.}
	\end{center}
\end{figure}

An example of a construction designed with five sine functions is shown in Fig.\ref{fig:Old_Sin_comparison}, in comparison with an example periodic function generated by a linear combination of Dirichlet kernels according to the method described in \cite{Tang-Garg-Kempf}. 
The superoscillations obtained from both methods have been designed to have the same zero crossings. Observe that with the new method, only a single lobe is present between each of these stretches, that is, every zero crossing lies within the superoscillation. 
By reading off the maximum amplitudes of the lobes and the superoscillations for both cases, we can conclude that the dynamic range does not differ here by more than a factor $\mathcal{O}(1)$: $\bar{S}(t)$ has slightly better dynamic range, a fact to be expected since it is in fact the $L^2$ minimizing curve. Let us recall, however, that the additive method is numerically unstable, in contrast to the new method. 

This example has the added benefit that we may decompose the product representation into a linear form, such as that shown below for N = 3 and $\epsilon_i = a$: 
\begin{equation}
S_3(t) = \frac{1}{4}\left(\sin\left(\frac{\Omega}{3}(t+2a)\right)+2\sin\left(\frac{\Omega}{3} t\right)-\sin\left(3\frac{\Omega}{3} t\right)\right)
\end{equation}
Note that each of the component oscillations have rational coefficients, which do not need to be evaluated to high precision to prevent the superoscillations from being destroyed by numerical error. 

\subsection{Generation of square-integrable superoscillating functions}
\begin{figure}[!ht]
	\begin{center}
		\includegraphics[width=0.8\columnwidth]{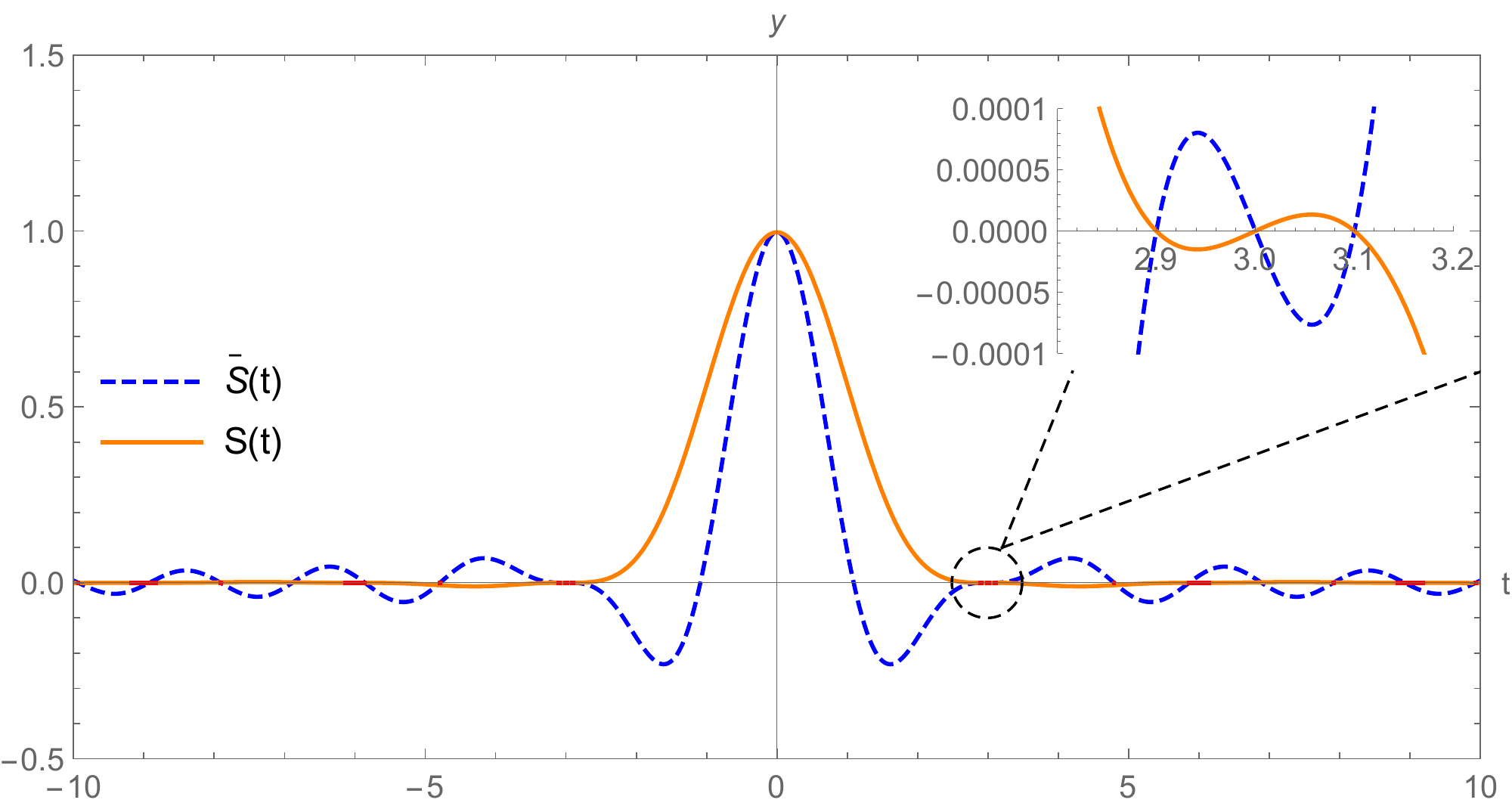}
		\caption{\label{fig:Old_Sinc_comparison} Two superoscillating curves $S(t)$ and $\bar{S}(t)$, obtained by the new multiplicative and the conventional additive method respectively.}
	\end{center}
\end{figure}

\begin{figure}[!t]
	\begin{center}
		\includegraphics[width=0.8\columnwidth]{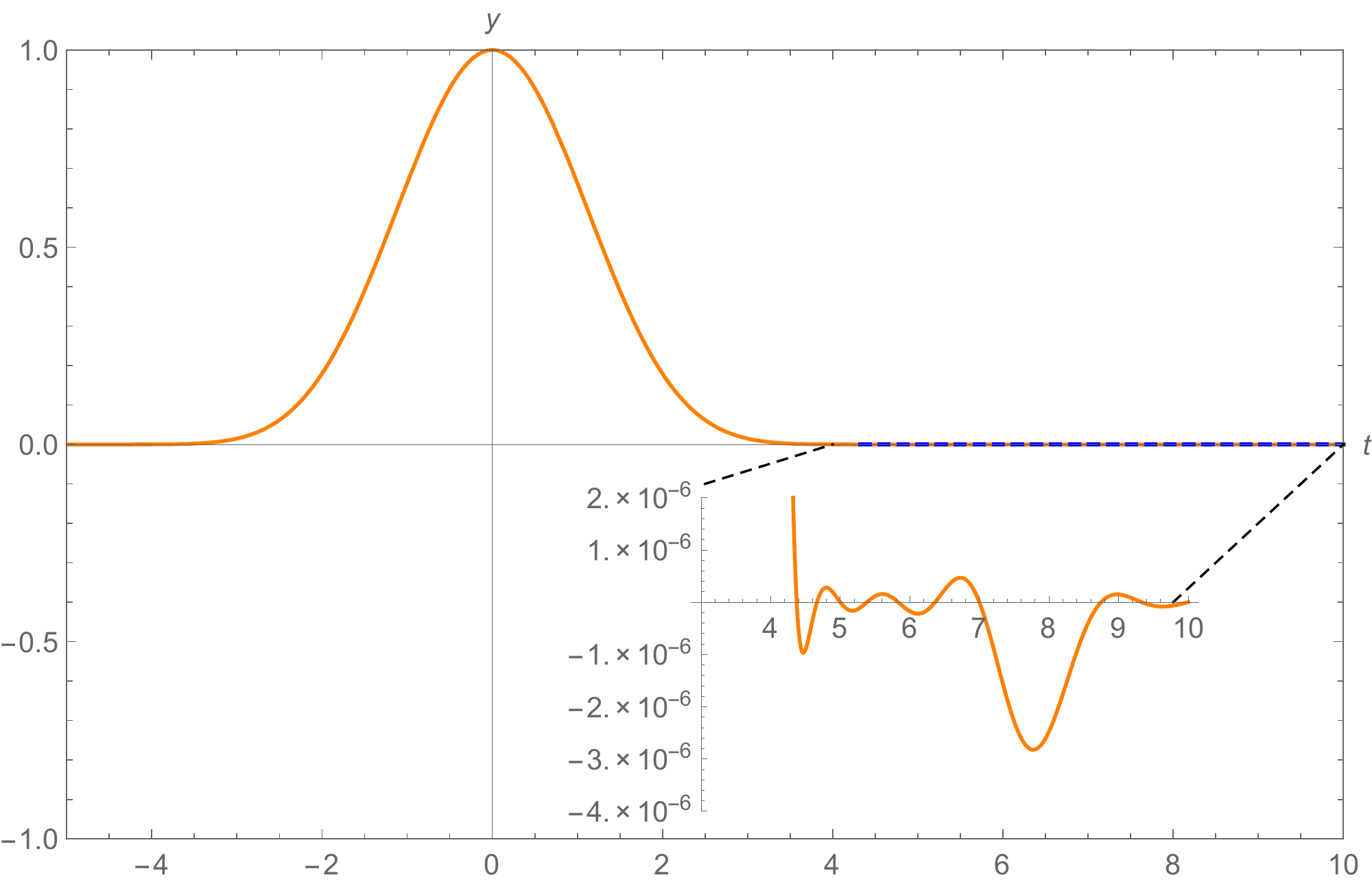}
		\caption{\label{fig:diff_BL} A superoscillating function constructed from multiplication of seven sinc functions of different bandlimits, all centred on the origin.}
	\end{center}
\end{figure}

Consider the bandlimited sinc function as an example of a square integrable factor, defined as:
\begin{equation}
\textrm{sinc}(\pi t) = \frac{\sin(\pi t)}{\pi t}
\end{equation}
It has real zeroes occurring at every integer value of $t$. Firstly, we may construct a superoscillating function by considering products of translated sinc functions, with relative displacements $\epsilon_i$. By, for example, choosing all the factor functions to have the same bandlimit we obtain the following construction:
\begin{equation}
S_N(t)=\prod_{n=1}^{N} \textrm{sinc}\left(\frac{\pi\Omega}{N} (t-\epsilon_n)\right)
\end{equation}
A specific example of this kind of function is shown in Fig.\ref{fig:Old_Sinc_comparison} for $\Omega = \pi$, $N=3$, and $\epsilon = 0.1$ corresponding to a superoscillating frequency of $\omega =10\pi$, in comparison with a function generated by the additive method described in \cite{Kempf-Ferreira-Unusual}. 
Examination of the figure shows there are several superoscillating stretches that arise regularly for free with the new method due to the periodicity of the sine component in the sinc function; however the superoscillations further out are decaying asymptotically as $1/x^N$. The locations of the prescribed zeroes are the same by design, as are also their maximum amplitudes, which occur at $t=0$.
Observe also that only a single lobe is present between each of these stretches since each zero crossing lies within a superoscillating stretch, whereas $\bar{S}(t)$ generated with the older approach presents several lobes between its two prescribed superoscillating stretches, which in this case results in its superoscillations being out of phase with those of $S(t)$. 
Further, we observe there is only a factor $\mathcal{O}(1)$ difference in the dynamic range between both methods. Again, we observe that the advantage of the new method does not come at a significant cost.
Secondly, aside from translating the factor functions, superoscillations may be generated by multiplying factors whose bandlimits are not equal, such that some oscillate more slowly than others and are stretched out more. Even the fastest of these must oscillate more slowly than dictated by the global desired bandlimit, such that the sum of individual bandlimits does not exceed the total limit. In fact, diversity in individual bandlimits alone is sufficient to generate superoscillating stretches as is illustrated in Fig.\ref{fig:diff_BL}. In this example, notice the irregularity of the shape and spacing of the superoscillations. The irregularly shaped superoscillating stretches obtained using this method may overlap, as in Fig.\ref{fig:diff_BL}, to combine into a single stretch for sufficiently large N and a given ratio of bandlimit to superoscillating frequency. Also note how the slowed factors can be arranged such that aside from the central lobe, the remainder of the function is superoscillatory.
In the above examples we have considered functions made from products of the same exclusively translated or stretched functions. $b(t)$. In general, combinations of these methods can be used to obtain superoscillating stretches of arbitrary length -by increasing the number of factor functions- and any desired local superoscillatory frequency determined by the zero spacings - in turn determined by choice of each $\epsilon_i$ and the relative bandwidths $\omega_i$. 

\subsection{Scaling properties of the dynamic range} 
We have so far qualitatively established that the dynamic ranges of the conventional additive and new multiplicative approaches are comparable for small $N$. 
It is of further interest to compare the scaling of the dynamic range $\sigma$ of these superoscillating functions against that of the conventional functions (which are energy minimizing solutions), in order to verify the conclusion that the dynamic range is on the same order of magnitude for both methods. To this end, upper and lower bounds on the dynamic range are established for each of the example constructions by identifying underestimates and overestimates for the maximum amplitudes of the lobes and of the superoscillations. 

\begin{figure}[!t]
	\begin{center}
		\includegraphics[width=0.8\columnwidth]{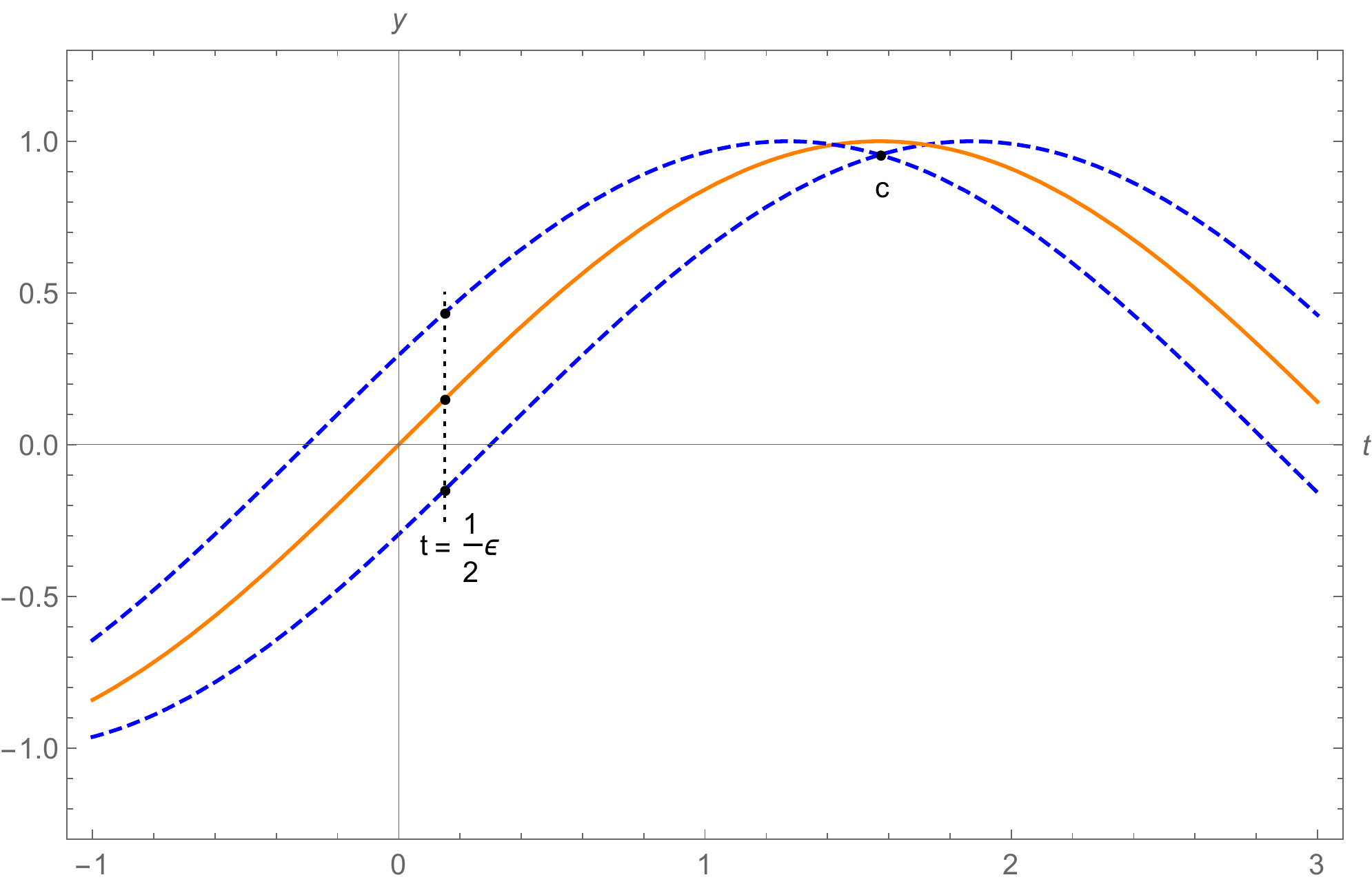}
		\caption{\label{fig:Sin_scaling} Three sinusoidal factor functions, whose product generates a superoscillating function. In dashed lines are the maximally translated factors, the solid line is a sine with no phase shift.}
	\end{center}
\end{figure}

Consider the undisplaced sin and maximally translated sin (displaced by a phase of $\epsilon(N-1)/2$) as shown in Fig.\ref{fig:Sin_scaling} .
An underestimate for the amplitude of the lobe can be obtained by taking the Nth power of the value of the translated sin at the location of the maxima of the origin centred sin:
\begin{equation}
c^N = \sin^N\left(\frac{\omega\pi}{N}\left(\frac{N}{2}+\frac{N-1}{2}\epsilon\right)\right)
= \cos^N\left(\frac{(N-1)\omega\pi\epsilon}{N}\right)
\end{equation}
Now consider the value of S(t) at the location halfway between two superoscillating zeroes $t = \epsilon/2$, which is given by the product of values of each sinusoid factor at $t = \epsilon/2$; each of these values lie on the dashed vertical line shown in Fig.\ref{fig:Sin_scaling}. Explicitly for a superoscillating construction made for example with an odd number of components N:
\begin{eqnarray}
\label{eq:c}
S_N\left(\frac{\epsilon}{2}\right) = \sin\left(\frac{\pi\omega\epsilon}{2N}\right)\sin\left(\frac{-\pi\omega\epsilon}{2N}\right)\sin\left(\frac{3\pi\omega\epsilon}{2N}\right)...\sin\left(\frac{\pi\omega\epsilon}{2}\right) \nonumber \\
  = \sin\left(\frac{\pi\omega\epsilon}{2}\right)\prod_{n=1}^{\frac{N-1}{2}}(-1)^{\frac{(N+1)}{2}} \left(\sin\left(\frac{n\pi\omega\epsilon}{2N}\right)\right)^2
\end{eqnarray}
We may disregard the negative factor since it is the magnitude and not the sign of the corresponding amplitude which is important. The relevant overestimate can be made by replacing each element in the product with the largest of these values:
\begin{equation}
S_N\left(\frac{\epsilon}{2}\right) \leq \left( \sin\left(\frac{\pi\omega\epsilon (N-1)}{2N}\right)\right)^N : = d
\end{equation}
The ratio of the two prior estimates will provide a lower bound on the dynamic range:
\begin{equation}
\sigma \geq \left(\frac{c}{d}\right)^N = \left(\frac{\cos\left(\frac{(N-1)\omega\pi\epsilon}{N}\right)}{ \sin\left(\frac{\pi\omega\epsilon (N-1)}{4N}\right)}\right)^N
\end{equation}
For large enough $N$, we have $\frac{N-1}{N}\sim 1$ and therefore:
\begin{equation}
\sigma \geq \left(\frac{\cos\left(\omega\pi\epsilon\right)}{\sin\left(\frac{\omega\pi\epsilon}{4}\right)}\right)^N
\approx \left(\frac{1-(\pi\epsilon)^2+\mathcal{O}(\epsilon^4)}{\frac{\pi\epsilon}{4}+\mathcal{O}(\epsilon^3)}\right)^N
\end{equation}
An upper bound may also be established by observing that $S_N(t) \leq 1$ for all $t$, with equality only satisfied if $N=1$, such that unity provides a suitable overestimate of the lobe amplitudes.
An underestimate of the superoscillatory amplitudes may be obtained by replacing the elements of the product in Eq. \ref{eq:c} with the Nth power of the smallest element: 
\begin{equation}
S_N\left(\frac{\epsilon}{2}\right) \geq \left(\sin \left(\frac{\omega\pi\epsilon}{2}\right) \right)^N : = b
\end{equation}
In this way we find: 
\begin{equation}
\sigma \leq \frac{1}{b^N} = \frac{1}{\left(\sin \left(\frac{\omega\pi\epsilon}{2}\right)\right)^N}
\end{equation}
The dynamic range is upper and lower bounded by quantities which scale polynomially in $\epsilon$ and exponentially in $N$. Therefore the dynamic range must scale in the same way.

We repeated a similar kind of analysis for the product of translated Sinc functions, revealing the following estimate:
\begin{equation}
\sigma \geq \left(\frac{c^{\prime}}{d^{\prime}}\right)^N = \left(\frac{\textrm{sinc}\left(\frac{\pi\epsilon(N-1)}{2N}\right)}{\textrm{sinc}\left(\pi-\pi\frac{N-1}{N}\epsilon\right)}\right)^N
\end{equation}
which when considering a large number of superoscillations $N$ reduces to 
\begin{equation}
\sigma \geq \left(\frac{\textrm{sinc}\left(\frac{\pi\epsilon}{2}\right)}{\textrm{sinc}(\pi(1-\epsilon))}\right)^N
\end{equation}
An upper bound may also be found with a form polynomial in $\epsilon$ and exponential in $N$. In fact, this kind of analysis may be carried out for any multiplicatively constructed superoscillating functions by taking the Nth power of upper and lower estimates of the values of the factor functions, and will always result in scaling that is exponential in $N$ and polynomial in $\epsilon$ since each individual bandlimited factor is entire and admits a polynomial Taylor representation.

The bounds here established for the new methods match the dynamic range scaling determined \cite{Kempf-Ferreira-Unusual} for the superoscillations generated by previous methods: the scaling is polynomial in the parameter $\epsilon$, hence polynomial upon increasing the superoscillating frequency $\nu$, and exponential upon increasing number of superoscillations N. Therefore, the quality of the superoscillations generated multiplicatively is comparable to those generated additively, which are additionally more immune to numerical difficulties; therefore, there is a decided advantage to the multiplicative method -namely the ease of mathematical implementation- {with little disadvantage, subject to possible limitations in the physical multiplication of functions. For example, while multiplication is difficult to achieve with light, the multiplication of electrical signals is straightforward. Also, in the case of periodic superoscillations, the need for implementing multiplication can be avoided altogether. This is because trigonometric identities can be used to give the periodic superoscillations a linear sum expansion. This, in turn, means that these superoscillating functions can be obtained by linear superposition.}

\section{New method for constructing superoscillating quantum systems}

\label{sec:wavefunction}
Let us now consider how Hamiltonians may be constructed which possess at least one superoscillatory eigenstate, as this could provide an avenue for physical implementations. 

To this end, let us start with an arbitrary wave function and use the Schr\"odinger equation to reverse engineer a Schr{\"o}dinger potential for it: 
\begin{equation}
V(x) = E_n + \frac{\hbar^2}{2m} \, \frac{\frac{d^2\Psi (x))}{d^2x}}{\Psi (x)}\;
\end{equation}
By redefining the energy scale such that $E_n = 0$ and absorbing the factor $\hbar^2/2m$ into the potential, we can simplify our expression for the potential:
\begin{equation} \label{eq:potential}
V(x) = \frac{\frac{d^2\Psi (x))}{d^2x}}{\Psi (x)}
\end{equation}
While zero-crossings can play an important role in superoscillatory functions, we read off from Eq.\ref{eq:potential}, that wherever the wave function crosses zero the potential may become singular. In fact, a change in sign of the wave function when zero is crossed can lead to singularities in the potential which will diverge to positive and negative infinity depending on the direction of approach. A potential with this kind of divergence is generally unphysical.

To avoid this problem one could try to ensure that the second derivative has zeroes at the same locations as the wave function; however, such an approach would limit the space of functions one may wish to consider.

Instead, let us remove the singularities in the potential by 
adding a constant term $C$ to the wave function, where $C$ can be positive or negative. 
The corresponding Schr\"odinger potential changes as we increase the lift C, having the form:
\begin{equation}
V(x) = \frac{\frac{d^2\Psi (x))}{d^2x}}{\Psi (x)+C}
\end{equation}
This procedure leaves the superoscillations intact and at the same time, for sufficiently large $|C|$, it removes the zeroes from the real line. In order to maintain normalizability, we will restrict attention to wave functions on a finite interval with periodic boundary conditions. 

We will now show that the resulting lifted wave function is the ground state of the so-obtained Hamiltonian: Sturm's separation and comparison theorems \cite{Sturm} state that solutions of a Sturm-Liouville problem, such as the Schr\"odinger equation, necessarily have an increasing number of zeroes. Furthermore, in order to ensure orthogonality, the zeroes must be placed where the wave function crosses the axis, not just touches it.
This means we can rule out the possibility of finding an eigenstate in the spectrum with a zero touching. All other eigenstates will have zero crossings, and intuitively must therefore have greater curvature and will therefore be higher energy states. Concretely, it was shown by Feynman \cite{Feynman} that any wave function with zero crossing cannot be the ground state. By elimination, this ensures that the lifted superoscillating wave function is the ground state.

\begin{figure}[!t] 
	\begin{subfigure}[b]{0.33\linewidth}
		\centering
		\includegraphics[width=1\linewidth]{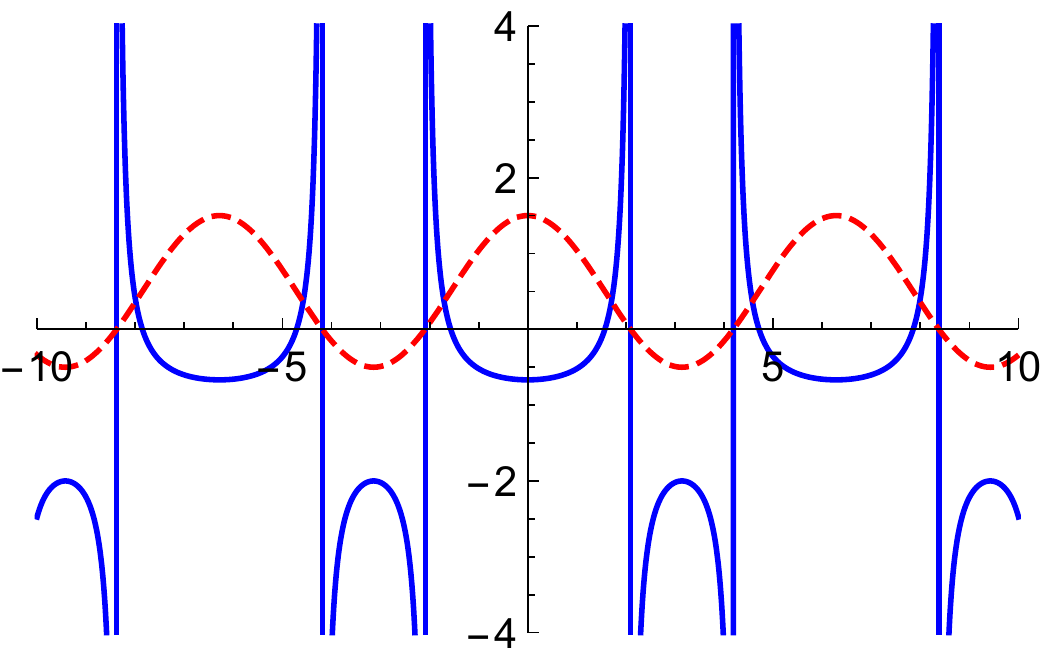} 
		\caption{Generic oscillatory function}
		\label{a}
	\end{subfigure} 
	\begin{subfigure}[b]{0.33\linewidth}
		\centering
		\includegraphics[width=1\linewidth]{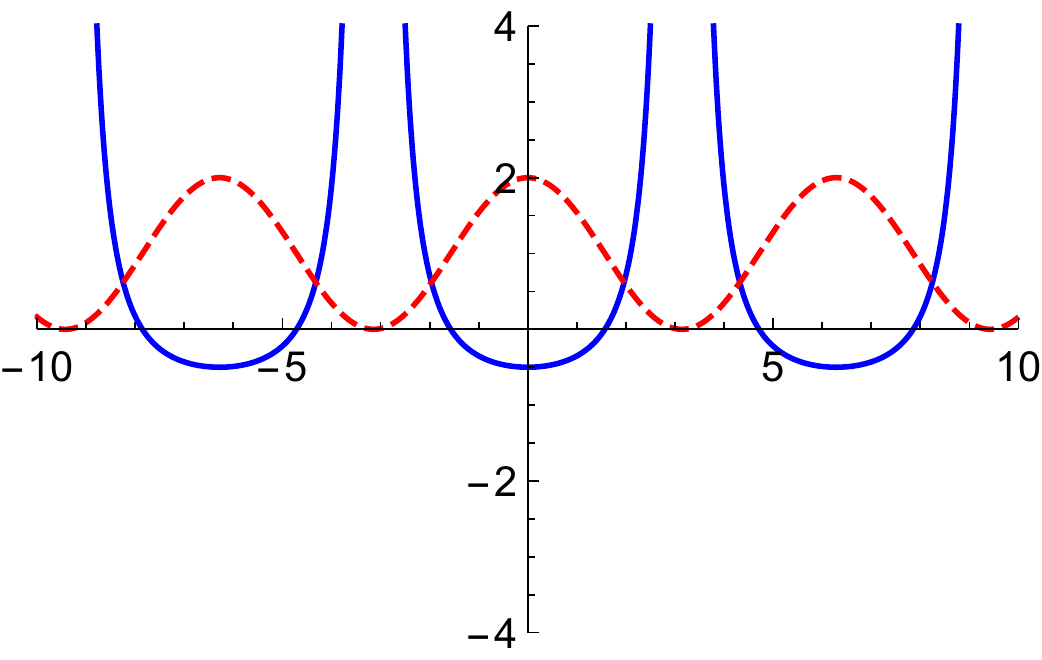} 
		\caption{Critical lift}
		\label{b}
	\end{subfigure}
	\begin{subfigure}[b]{0.33\linewidth}
		\centering
		\includegraphics[width=1\linewidth]{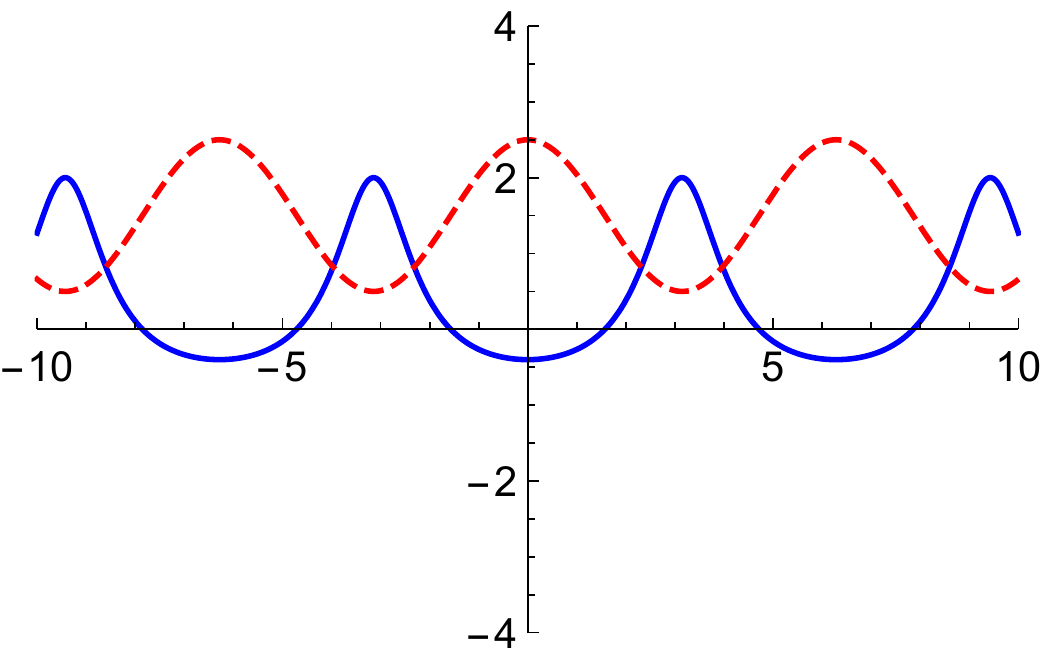} 
		\caption{Sufficient lift}
		\label{c}
	\end{subfigure} 
	\caption{
	{Each image displays -as an example- a lifted sinusoidal wave function (dashed line), and its corresponding Schr\"odinger potential (solid lines). Observe that the potential loses its singularities once the lift is equal to or in excess of the most negative value of the original wave function.} 
	}
	\label{fig:Lifting_zeroes} 
\end{figure}

Let us now return to how the process of lifting affects the occurrence of singularities in the potential. For example, consider the case of a sinusoidal wave function lifted by varying amounts. Fig.\ref{fig:Lifting_zeroes} shows how the potential changes as $C$ is varied.   

Fig.\ref{fig:Lifting_zeroes}\subref{a} displays the case of no lift in which the zero crossings induce divergences in the reverse-engineered potential whose left and right hand limits differ by a sign and are therefore unphysical.  
Fig.\ref{fig:Lifting_zeroes}\subref{b} displays the case of a critical lift (with at most zero touchings), where the potential possesses more realistic divergences since they do not change sign. Fig.\ref{fig:Lifting_zeroes}\subref{c} displays the case of sufficient lift (no zero crossings), in which the potential no longer has divergences. 

\begin{figure}[!t] 
	\begin{subfigure}[b]{0.5\linewidth}
		\centering
		\includegraphics[width=1\linewidth]{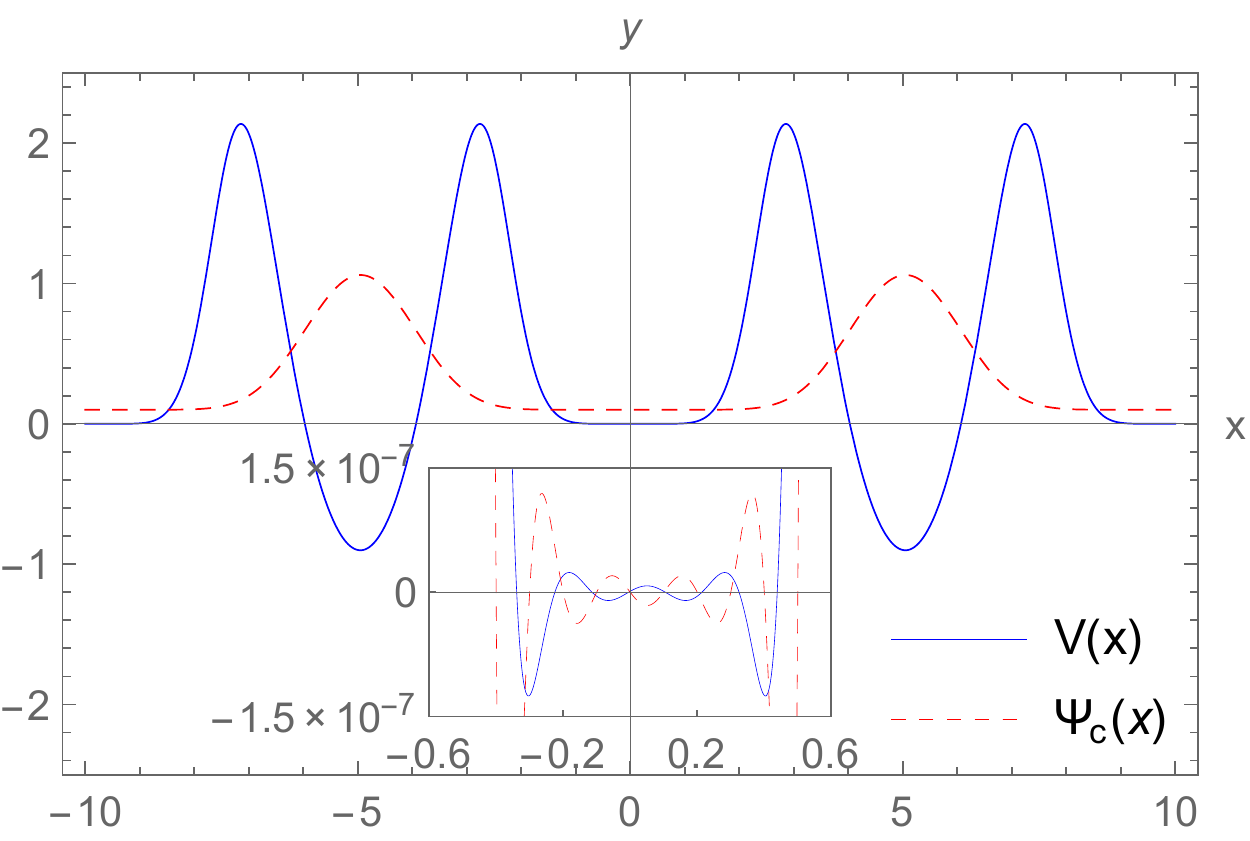} 
		\caption{Positive lift}  
		\label{fig:Potentials-Pos}
		\vspace{4ex}
	\end{subfigure}
	\begin{subfigure}[b]{0.5\linewidth}
		\centering
		\includegraphics[width=1\linewidth]{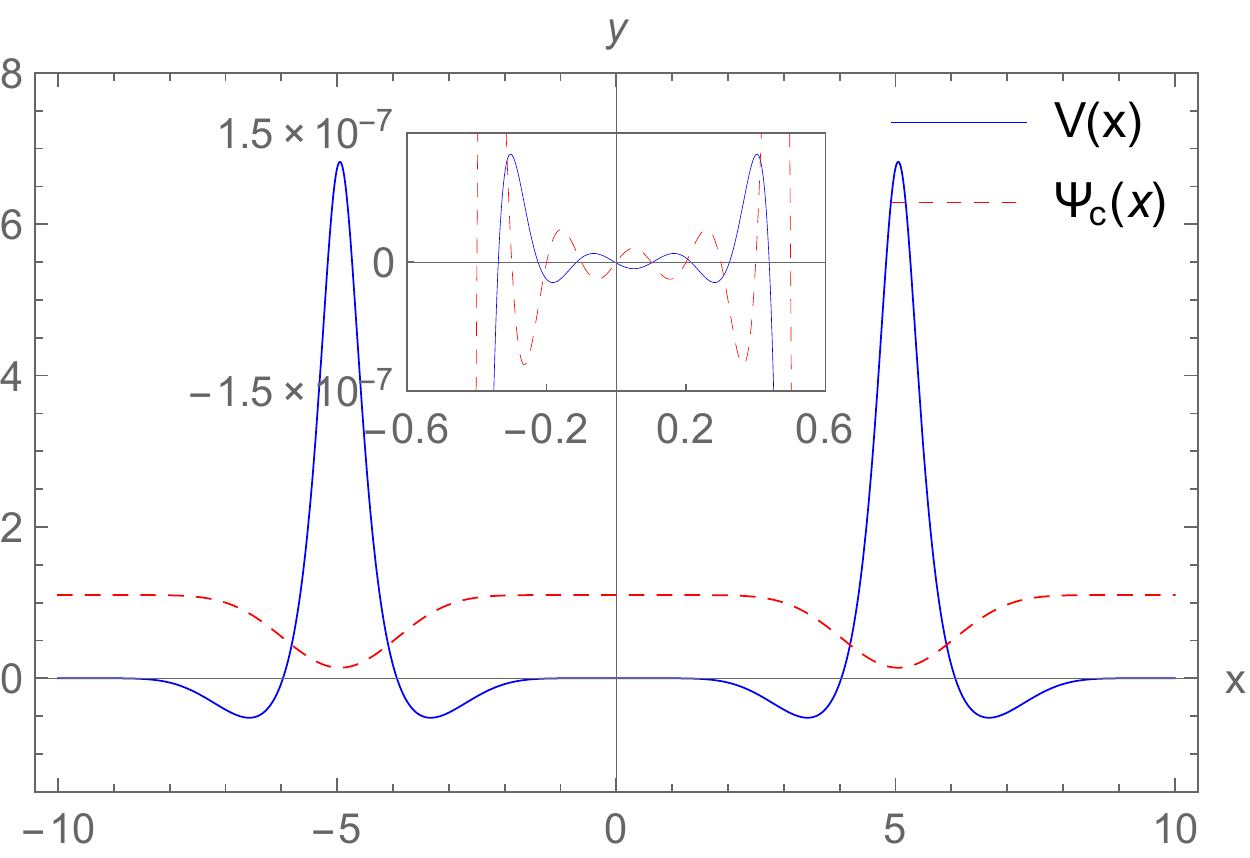} 
		\caption{Negative lift}  
		\label{fig:Potentials-Neg}
		\vspace{4ex}
	\end{subfigure} 
	\caption{The potentials evaluated for different lifted superoscillating wave functions. Overlaid is the corresponding ground state eigenfunction (dashed line), in arbitrary units.}
	\label{fig:Potentials} 
\end{figure}

Fig.\ref{fig:Potentials} shows two Schr\"odinger potentials generated in this way for a concrete periodic superoscillating wave function. 
Examination of the insets reveal that the potential is highly oscillatory in the regions corresponding to the superoscillating stretch. This appears to be a general feature of potentials that generate superoscillating eigenstates. 

It is of interest next to interpret the distribution of probability density for each of these cases. Let us first consider the case of positive lift: in Fig.\ref{fig:Potentials}\subref{fig:Potentials-Pos} we see that the two high probability amplitude lobes of the wave function line up with two deep wells in the potential. This is to be expected, since here the energy is lowest. 

Each of these wells in the potential is flanked by two large energy barriers, which the particle may tunnel through to arrive at the superoscillating plateau in the center of the graph, where the potential energy is higher than in the well. In this way, it is less probable to find the particle in the superoscillating region, as one would expect given the large dynamic range of the original superoscillating construction. If we increase the lift, this decreases the height of the energy barriers and the depth of the wells, which in turn tends to equalize the spread of probability amplitudes as the particle has smaller energy differences to overcome. 


We now consider the case of negative lift, an example of which is shown in Fig.\ref{fig:Potentials}\subref{fig:Potentials-Neg}. In this case, $|C|$ must be chosen larger than the peak of the lobes. As the figure shows, the wave function then has large probability amplitudes in the superoscillating plateau between the peaks of the potential and, curiously, the wave function possesses relatively small probability amplitudes in the energy minima near the potential barriers. 

{In summary, the value of C must always be at least as large as the most negative value of the original wavefunction in order for all real line zeroes to be removed. In the case of negative lift, C is at most on the order of unity (for a normalized wavefunction). It is typically quite small in the case of positive lift, for which a superoscillating wavefunction is lifted by an amount strictly greater than the largest superoscillating amplitude. This must be orders of magnitude smaller than unity, due to the exponential suppression of superoscillations.
The consequence of this is that potentials of the form obtained through positive lift will have large amplitudes in the non-superoscillating regions, whilst those obtained from negative lifts will have larger amplitudes in the superoscillating regions.  
In either case, a high spatial and energetic resolution must be achievable in order to construct such a potential. 
}

\section{Conclusions and Outlook}
We developed a new method for calculating bandlimited, superoscillatory functions $S(t)$ by multiplying functions $b(t)$ of suitably lower bandwidth. The fact that $S(t)$ inherits all zeroes of these functions allows one to design $S(t)$ to possess arbitrarily close zeroes and therefore superoscillations. There are no numerical difficulties in this construction, which means that functions with arbitrarily fast and long superoscillations can be composed straightforwardly. 

We also proved that the quality of the superoscillations produced in this way does not significantly suffer in comparison with older methods, in the sense that the dynamic range possesses the same scaling behavior as in the case of the construction that yields the superoscillatory functions with the smallest dynamic range possible (with the dynamic range defined in the $L^2$ norm sense, see \cite{Kempf-Ferreira-Unusual}). We note in passing that the method provides the option to generate the necessary large side lobes on only one side of the superoscillating stretch. 

We applied the multiplicative method to generate two physically particularly relevant cases of superoscillatory functions $S(t)$, namely functions that either possess a finite $L^2$ norm on the real line or that are periodic. 

While the new multiplicative approach provides a direct avenue for the mathematical generation of superoscillating functions, let us now briefly discuss how, analogously, it may be possible to physically create superoscillatory wave forms by using physical processes that in effect implement multiplication. For example, the multiplication of electronic signals is straightforward through the use of transistors. 
Furthermore, superoscillating functions constructed from products of purely trigonometric factor functions - which are therefore necessarily periodic - may be expanded into linear combinations using trigonometric product-sum identities. These constructions can therefore be implemented also by linearly combining monochromatic waves, for example, by using lasers. We here only remark that the new multiplicative approach to generating superoscillatory functions generalizes straightforwardly to multiple dimensions. 

We also presented a method for constructing Hamiltonians which possess nonsingular potentials and which possess the property that an eigenstate possesses a superoscillatory wave function. In particular, we showed that a Hamiltonian can always be found such that any chosen periodic superoscillating wave function, suitably lifted, is the ground state. It remains an interesting open question to what extent the wave functions of the excited states then may also possess superoscillatory behavior. 
$$$$
\noindent \bf Acknowledgement: \rm AK acknowledges support under the Discovery program of the National Science and Engineering Research Council (NSERC) of Canada.

\end{document}